\documentclass{article}

\usepackage{amsmath,amssymb,mathtools} 
\usepackage{bm}                        
\usepackage[english]{babel}
\usepackage{float}  
\usepackage[letterpaper,top=2cm,bottom=2cm,left=3cm,right=3cm,marginparwidth=1.75cm]{geometry}
\usepackage{amsmath}
\usepackage{amssymb}
\usepackage{multirow}
\usepackage{booktabs} 
\usepackage{amsmath}
\usepackage{graphicx}
\usepackage{mathtools}  
\usepackage[colorlinks=true, allcolors=blue]{hyperref}
\usepackage{natbib}
\usepackage{authblk} 
\title{DiffTopo: Solver in the Loop for Inverse Topography via Condition Diffusion Generation}
\author[1,2]{Aoming Liang}
\author[2]{Qi Liu}
\author[2]{Weicheng Cui}
\affil[1]{College of Environmental and Resource Sciences, Zhejiang University}
\affil[2]{School of Engineering, Westlake University}

\begin{document}
\date{} 
\maketitle

\begin{abstract}
Inferring seabed topography from wave‐height observations is fundamental to tsunami hazard assessment, coastal planning, and large-scale ocean circulation modeling. Classical inversion models typically rely on direct-sensing or optimization-based schemes that must contend with the strongly nonlinear coupling between free-surface dynamics and topography. However, data-driven approaches are capable of tackling strongly nonlinear problems by learning the underlying data distributions. This study introduces DiffTopo, a conditional diffusion model that reconstructs topography from surface wave field data governed by shallow-water equations. Using classifier-free guidance, DiffTopo not only generates a series of solutions but also applies a thresholding mechanism that ensures, via the solver, the validation results are physically plausible. This study evaluates both observed wave fields and three distinct topography configurations, demonstrating that DiffTopo exhibits robust generalization and remains consistent with the shallow water equations even under full observations. These results underscore the potential of diffusion-based generative modeling for addressing ill-posed inverse problems in geophysics.

\end{abstract}

\section{Introduction}

In numerous scientific and engineering domains, an inverse problem involves deducing the underlying causes or model parameters of a system from observed measurements \citep{tarantola2005inverse}. This stands in contrast to the forward problem, which aims to predict system responses given a known set of inputs. Inverse problems are inherently challenging because of their ill-posed nature, which lacks a unique or stable solution. In the field of ocean engineering, the forward problem is formulated as predicting the temporal evolution of waves given the initial wave conditions and the geometric characteristics of the seabed topography \citep{mu2025generalizing}. In contrast, the inverse problem involves inferring the underlying topography from observations of wave evolution over time \citep{holman2013cbathy}.

Accurate knowledge of ocean bathymetry is critical to ensuring safe underwater navigation \cite{vasan2013inverse} and water resources management. The topography of the seafloor plays a fundamental role in the regulation of water movement \citep{anderson1978role,flament2013review}. Variations in seabed elevation critically influence the behavior of surface waves \citep{snieder1988large}, tides \citep{egbert1997tidal}, tsunamis \citep{melgar2015kinematic}. For example, changes in water depth cause surface waves to refract, reflect, or diffract, directly shaping nearshore wave patterns. The geometry of the seabed determines the flow pathways and intensities of tidal currents. Moreover, during tsunami events, shallower topography near coastlines can significantly amplify wave heights and alter their propagation, with potentially devastating consequences for coastal engineering \citep{li2019wave}. A robust topography inversion model is essential for optimizing infrastructure design \citep{narayanan2004model}.

Methods for inferring seafloor topography from surface wave elevations primarily include physics-based inversion, e.g., adjoint-based data assimilation \citep{wu2023adjoint} or variational optimization to reconstruct seafloor depth from observed wave transformations \citep{desmars2023nonlinear} and system identification \citep{liang2023system}.  High-resolution imagery from satellites to extract wave kinematics and infer nearshore topography via dispersion analysis or a deep convolutional network \cite{xi2025band,sun2025nearshore}. Recent advances have explored physics-driven inversion methods  \citep{fanous2025leveraging} and data-driven methods \citep{kabiri2025depth} to recover topography from wave observations. However, many of these approaches either assume full observability or require expensive PDE-constrained optimization methods \citep{angel2024bathymetry}. Neural operator-based methods often struggle to produce accurate results when solving high-dimensional inverse problems \citep{liang2024mixed,wang2024latent}. \citet{liu2024bathymetry} proposes the CNN autoencoder to reconstruct the 2D river bathmetry.

The recent success of generative models has brought new hope to solving inverse problems. Due to the ill-posed nature of many inverse problems, traditional optimization-based approaches often struggle to recover the optimal solution, especially when the solution space is high-dimensional or underconstrained. Generative models can provide a data-driven alternative by learning the underlying distribution of plausible solutions, thereby enabling more robust and realistic reconstructions. \citet{farimani2017deep} presents a condition GAN to solve the non-linear transport equations. \citet{huang2024diffusionpde} proposes the DiffusionPDE, which can simultaneously fill in missing information by modeling the joint distribution of the solution and coefficient spaces. \cite{shysheya2024conditional} introduces a comparative study that is conducted on score-based diffusion models for prediction and assimilation with sparse observational data. 
\citet{haitsiukevich2024diffusion} suggests a mix-condition diffusion model that trains a single model capable of adapting to multiple tasks by alternately performing different tasks during the training process. \citet{li2025generative} develops a generative solver to estimate the inverse problem by latent flow matching. \citep{wang2024wavediffusion} used the latent diffusion method to incorporate seismic data and velocity data to reconstruct the seismic waveform field.

Although studies on diffusion models are growing, their application to ocean wave dynamics and topography inversion remains limited. This work explores the potential of diffusion models in this context, with a particular focus on classifier-free guidance for conditional generation. Traditional conditional diffusion models, which depend on the classifier guide \citep{hu2023self}, require training an additional noise-resistant classifier to steer the generation process. However, this becomes challenging when the conditioning input is a complex, high-dimensional continuous field, such as a spatio-temporal wave elevation. In contrast, classifier-free guidance (CFG) \citep{tang2025diffusion} offers a flexible and unified approach, enabling smooth interpolation between unconditional and strongly conditional generation by adjusting a single parameter of the guidance scale. This makes CFG particularly well suited for our task of topography inversion from sparse wave observations, where controlling the influence of the wave field on the generation process is critical.

In this work, we introduce DiffTopo, a diffusion-based method for generating topography models by learning the posterior distribution of the topography from observed water waves. We validate DiffTopo on different topologies by posterior validation, demonstrating its ability to reconstruct three common topography patterns. DiffTopo highlights the potential of diffusion-based inverse modeling as a robust and generalizable solution for ocean applications.

\section{Methodology}

 The underlying principles of the dataset and the evaluation objectives are as follows.

\subsection{Shallow Water Equations and numerical solver setting}
We consider nonlinear wave propagation over a static seafloor in a two-dimensional spatial domain
\(\Omega = [0,L_x] \times [0,L_y] \subset \mathbb{R}^2\) for \(t \in [0,T]\). 
The governing equations are the nonlinear shallow-water equations (SWEs) with Manning friction in the reference \citep{sanders2000adjoint,leveque2002finite}.
\begin{equation}
\begin{cases}
\displaystyle 
\frac{\partial \eta}{\partial t} 
+ \frac{\partial M}{\partial x} 
+ \frac{\partial N}{\partial y} = 0, \\[1.5ex]
\displaystyle 
\frac{\partial M}{\partial t} 
+ \frac{\partial}{\partial x} \left( \frac{M^2}{D} + \frac{1}{2} g D^2 \right) 
+ \frac{\partial}{\partial y} \left( \frac{MN}{D} \right) 
= -\,gD\,\frac{\partial h}{\partial x} 
- g \alpha^2\,\frac{M \sqrt{M^2 + N^2}}{D^{7/3}}, \\[1.5ex]
\displaystyle 
\frac{\partial N}{\partial t} 
+ \frac{\partial}{\partial x} \left( \frac{MN}{D} \right) 
+ \frac{\partial}{\partial y} \left( \frac{N^2}{D} + \frac{1}{2} g D^2 \right) 
= -\,gD\,\frac{\partial h}{\partial y} 
- g \alpha^2\,\frac{N \sqrt{M^2 + N^2}}{D^{7/3}},
\end{cases}
\label{eq:swe}
\end{equation}
where:
\begin{itemize}
    \item $g = 9.81~\mathrm{m/s^2}$: gravitational acceleration.
    \item $\alpha = 0.025~\mathrm{m^{-1/3} \cdot s}$: Manning roughness.
    \item $h(x,y)~[\mathrm{m}]$: still-water topography depth.
    \item $\eta(x,y,t)~[\mathrm{m}]$: free-surface elevation relative to still water level.
    \item $D(x,y,t)~[\mathrm{m}]$: total height, $D = h + \eta$.
    \item $u, v~[\mathrm{m/s}]$: depth-averaged horizontal velocity components.
    \item $M~[\mathrm{m^2/s}]$: depth-integrated momentum in the $x$ direction, $M = u D$.
    \item $N~[\mathrm{m^2/s}]$: depth-integrated momentum in the $y$ direction, $N = v D$.
\end{itemize}

\paragraph{Numerical discretization}
The computational domain is discretized on a uniform Cartesian grid with $N_x = N_y = 128$, $\Delta x = L_x / N_x$, $\Delta y = L_y / N_y$, where $L_x = L_y = 100~\mathrm{m}$.

\paragraph{Boundary condition}
The Neumann boundary conditions for the elevation of the free surface $\eta$ in a rectangular domain \(\Omega=[0,L_x]\times[0,L_y]\),
\begin{equation}
\left.
\frac{\partial \eta}{\partial x}\right|_{x=0}
=
\left.
\frac{\partial \eta}{\partial x}\right|_{x=L_x}
=0,
\qquad
\left.
\frac{\partial \eta}{\partial y}\right|_{y=0}
=
\left.
\frac{\partial \eta}{\partial y}\right|_{y=L_y}
=0.
\label{eq:eta-neumann}
\end{equation}

\paragraph{Initial condition}
A Gaussian pulse initializes the free-surface elevation:
\begin{equation}
\eta(x,y,0) = A \exp\!\left[
-\frac{(x-x_c)^2}{2\sigma_x^2} - \frac{(y-y_c)^2}{2\sigma_y^2}
\right],
\label{eq:ic-eta}
\end{equation}
with momenta \(M(x,y,0) = \kappa\,\eta(x,y,0)\) and \(N(x,y,0) = 0\). Other parameters are \((x_c, y_c) = (30,\, 50)~\mathrm{m}\), \(A = 0.5~\mathrm{m}\), \(\sigma_x = \sigma_y = \sqrt{2.5}~\mathrm{m}\), and \(\kappa = 100\).

\paragraph{Topography generation}
In this study, we investigate three common topographic configurations: single-seamount topographies, tanh-shaped topographies, and multi-seamount topographies. The corresponding parameterized formulations are presented in the following.

\begin{enumerate}
    \item \textbf{Single seamount Topography (SMT):}  
    A Gaussian spot of peak height $H_p$ and standard deviation $\sigma_h$ is superimposed on a uniform base depth $h_0$.  
    The spot center $(c_x, c_y)$ is randomly sampled within a restricted region  
    $c_x, c_y \in [20, 60]~\mathrm{m}$ to avoid proximity to the domain boundaries.  
    The topography is  given by:
    \begin{equation}
        h(x,y) = h_0 - H_p \exp\!\left( -\frac{(x-c_x)^2 + (y-c_y)^2}{2\sigma_h^2} \right).
    \end{equation}
    In our experiments, $h_0 = 30.0~\mathrm{m}$, $H_p = 20.0~\mathrm{m}$, and $\sigma_h = 8.0~\mathrm{m}$.

    \item \textbf{Tanh Topography (TanT):}  
A hyperbolic–tangent ridge is superimposed on a uniform base depth \(h_0\).  
The ridge center \((c_x,c_y)\) is uniformly sampled within an interior subdomain to avoid proximity to the domain boundaries, and the ridge orientation is randomized by an angle \(\theta \in [0,\pi)\).  
Let the rotated streamwise coordinate be
\begin{equation}
    \xi(x,y) = (x - c_x) \cos\theta + (y - c_y) \sin\theta,
\end{equation}
where \(H_r > 0\) denotes the ridge half-amplitude and \(s > 0\) controls the steepness of the slope (smaller \(s\) corresponds to steeper transitions).  $\xi(x,y)$ denotes the rotated streamwise coordinate obtained by translating the domain to $(c_x,c_y)$ and rotating by $\theta$.

The topography is calculated by:
\begin{equation}
    h(x,y) = h_0 + H_r \, \tanh\!\left( \frac{\xi(x,y)}{s} \right).
\end{equation}

with the maximum cross-ridge slope magnitude \(|\nabla h|_{\max} = H_r/s\) occurring at \(\xi = 0\).  
In our experiments, we set \(h_0 = 30.0~\mathrm{m}\), \(L_x = L_y = 100.0~\mathrm{m}\), \(H_r = 5.0~\mathrm{m}\), sample \((c_x,c_y) \sim \mathcal{U}(0.3L_x, 0.7L_x) \times \mathcal{U}(0.3L_y, 0.7L_y)\), \(\theta \sim \mathcal{U}(0,\pi)\), and \(s \sim \mathcal{U}(0.05L_x,\,0.20L_x)\).

    \item \textbf{Multi seamount Topography (MMT):}  
    Starting from a constant depth $h_0$, we add a zero-mean random perturbation field $r(x,y)$ uniformly sampled in $[-\rho, \rho]$,  
    with $\rho$ the perturbation amplitude, followed by a Gaussian smoothing kernel of width $\sigma_s$ to enforce spatial smoothness:
    \begin{equation}
        h(x,y) = h_0\,[1 + \tilde{r}(x,y)],
    \end{equation}
    where $\tilde{r}$ is the smoothed perturbation field.  $h_0 = 30.0~\mathrm{m}$, $\rho = 5.0$, and $\sigma_s = 8.0~\mathrm{m}$ in all random terrain cases.
\end{enumerate}

\paragraph{Simulation time and temporal sample for $\eta(x,y,t)$}
Each case is integrated over a total physical time of $T_{\mathrm{max}} = 6.0~\mathrm{s}$ with a uniform time step $\Delta t = \frac{1}{800}~\mathrm{s} = 1.25\times 10^{-3}~\mathrm{s}$, yielding $N_t = T_{\mathrm{max}} / \Delta t = 4800$ time steps per simulation by finite difference solver. For observation and storage efficiency, the elevation of the free surface $\eta$ is recorded at every $s = 100$ time step, resulting in stored frames $T = N_t / s = 48$ for each case. The observations $\eta_{ob}$ are arranged in a tensor of shape $[B,\, T,\, H,\, W]$, where $B$ is the number of cases, $T$ the number of frames stored, and $(H, W)$ the spatial resolution. Topography profiles are generated on a uniform grid of size $(n_x, n_y)$ in shape tensors $[B, H, W]$ for simulation. Since the finite-difference method is explicitly, the Courant–Friedrichs–Lewy number is maintained below 0.8 under this study.

\subsection{Principle: Conditional Generation with Classifier-Free Guidance}

 DiffTopo follows the standard Denoising Diffusion Probabilistic Model (DDPM \citep{ho2020denoising}) definition, with CFG to incorporate observation $\eta_{ob}$. Our approach consists of three main stages: training, inference sampling, and validation. In this study, the condition $c$ represents the observation $\eta_{ob}$. 
 The goal is to generate the topography $\hat{h}(x,y)$ based on observation $\eta_{\mathrm{ob}}$. 

\begin{equation}
\eta_{ob}(x, y, t) \in \mathbb{R}^{T \times H \times W}
\;\xmapsto{\text{DiffTopo}}\;
\hat{h}(x, y) \in \mathbb{R}^{H \times W}.
\end{equation}

\subsubsection{Training objective with classifier-free guidance}

Given a clean topography $x_0 \sim q(x_0)$, a noisy version $x_t$ in step $t$ is obtained by
\begin{equation}
x_t = \sqrt{\bar{\alpha}_t} x_0 + \sqrt{1 - \bar{\alpha}_t} \, \epsilon, 
\quad \epsilon \sim \mathcal{N}(0, \mathbf{I}),
\end{equation}
where $\bar{\alpha}_t = \prod_{s=1}^t \alpha_s$ and $\alpha_s = 1 - \beta_s$ 
follow the variance schedule $\{\beta_s\}_{s=1}^T$. Instead of directly modeling $p_\theta(x_{t-1} \mid x_t, c)$, the neural network $\epsilon_\theta$ is trained to predict the added noise $\epsilon$:
\[
\mathcal{L}_{\text{CFG}} = \mathbb{E}_{x_0, \epsilon, t, c} \left[ \| \epsilon - \epsilon_\theta(x_t, \tilde{c}) \|^2 \right],
\]
where the effective condition $\tilde{c} \in \{ c, \emptyset \}$ is chosen according to a Bernoulli distribution with a drop probability $p$ that is set to 0.1 in all experiments, following \cite{dhariwal2021diffusion}. The underlying principles of the conditional free guidance approach are detailed in the Appendix.

\subsubsection{Inference with Guidance and solver}

To improve conditional generation in the sampling process, the model is jointly trained with and without the condition \( c \). At inference time, we interpolate between the conditional and unconditional predictions using a guidance weight \( w > 0 \), and adjust the predicted noise as follows:
\begin{equation}
\hat{\epsilon}_\theta(x_t, c) = (1 + w) \cdot \epsilon_\theta(x_t, c) - w \cdot \epsilon_\theta(x_t, \varnothing),
\end{equation}
where \( \varnothing \) denotes the unconditional input (e.g., a zero tensor). This guided prediction \( \hat{\epsilon}_\theta \) is used in the reverse sampling of DDPM:
\begin{equation}
x_{t-1} = \frac{1}{\sqrt{\alpha_t}} \left( x_t - \frac{1 - \alpha_t}{\sqrt{1 - \bar{\alpha}_t}} \hat{\epsilon}_\theta(x_t, c) \right) + \sigma_t z,
\end{equation}
where \( z \sim \mathcal{N}(0, \mathbf{I}) \), and \( \sigma_t \) is the variance term determined by the noise schedule $\{\beta_s\}_{s=1}^T$.
In the inference stage, three representative sampling strategies were evaluated to assess generation quality and computational efficiency:
(1) the original DDPM sampler~\cite{ho2020denoising}, which follows the standard denoising 
diffusion probabilistic model formulation; 
(2) the Heun method~\cite{karras2022elucidating}, a second-order stochastic solver that improves
stability and accuracy during the reverse process; and
(3) the DPM++ sampler~\cite{lu2025dpm}, which leverages high-order integration techniques to
achieve faster convergence and better sample quality.

\begin{figure}[htbp]
    \centering
    \includegraphics[width=0.9\linewidth]{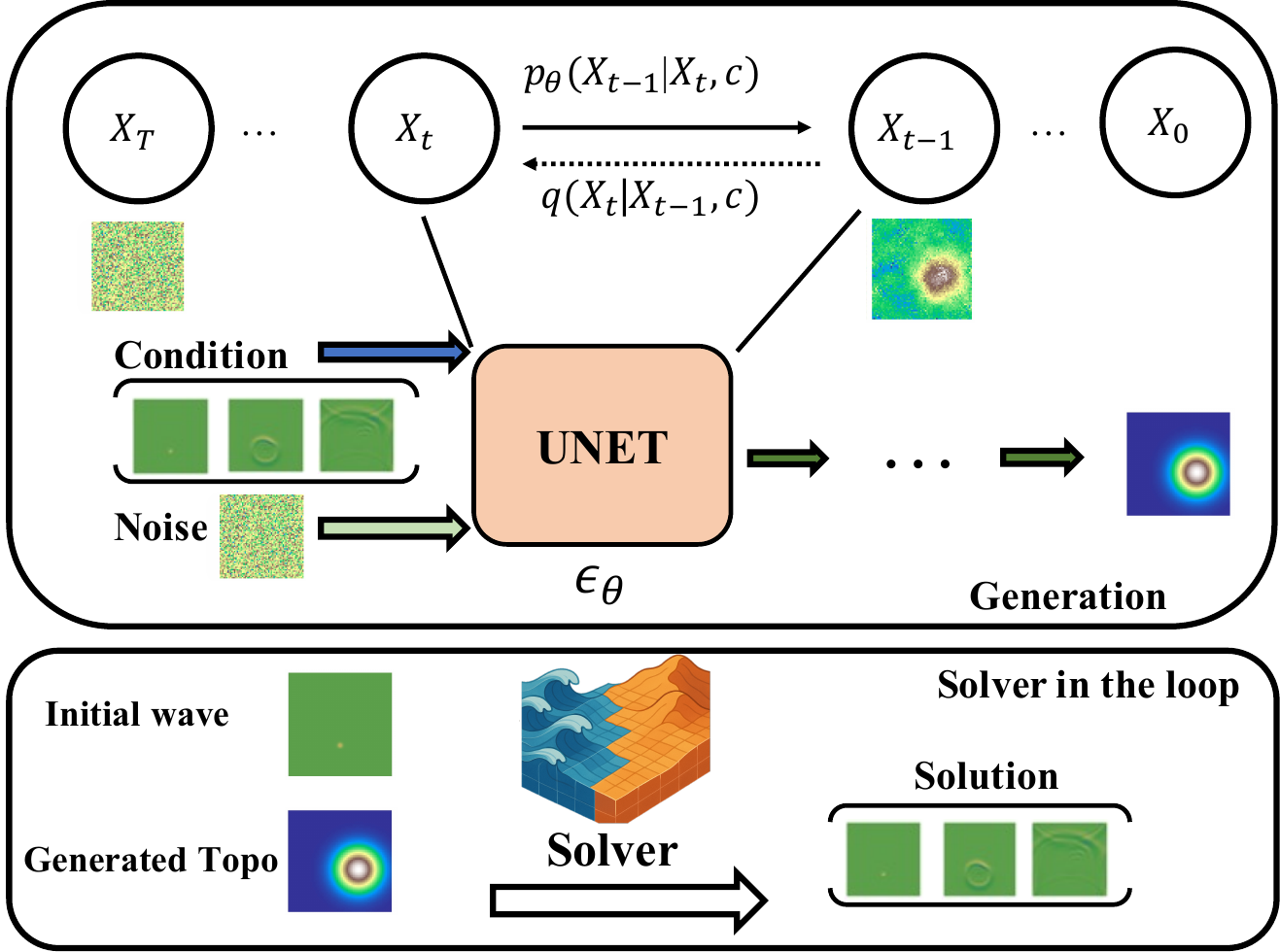}
    \caption{The upper part of the DiffTopo generation process illustrates the topography generation from wave field observations. The lower part displays the validation process, which verifies the topography on solver feedback until the residual is satisfied.}
    \label{fig:enter-label}
\end{figure}

\begin{table}
\centering
\caption{The settings and representation of three topographies. }
\begin{tabular}{c|c|c|c}
\hline
\textbf{Setting} & \textbf{Temporal steps (T)} & \textbf{Height (H)} & \textbf{Width (W)} \\
\hline
TanT& 48 & 128 & 128 \\
SMT& 48 & 128 & 128 \\
MMT & 48 & 128 & 128 \\
\hline
\end{tabular}
\label{tab:wave_resolutions}
\end{table}

\subsubsection{The posterior validation process }
Varying the guidance weight $\omega$ during the generation process results in significantly different generations. Given the generative nature of sampling, it is critical to assess the reliability of the results. In this section, we propose a posterior validation (solver in the loop) in which a numerical solver is used to regenerate the wave field from the generated topography. If $\eta^{valid}$ aligns with $\eta^{\mathrm{ob}}$, the corresponding topography is considered reliable.
To quantitatively assess the validation of the generated topography $\hat{h}$, DiffTopo enters it through the shallow-water solver to obtain the simulated wave field $\eta_{\mathrm{sim}}$. 
The mean squared error (MSE) between $\eta_{\mathrm{sim}}$ and
$\eta_{\mathrm{ob}}$:
\begin{equation}
\mathrm{R}(\hat h)
= \frac{1}{T H W}
\sum_{t=1}^{T}\sum_{i=1}^{H}\sum_{j=1}^{W}
\Big(\eta_{\mathrm{sim}}^{(t)}(i,j;\hat h)-\eta_{\mathrm{obs}}^{(t)}(i,j)\Big)^2 .
\end{equation}

This residual $\mathcal{R}$ serves as a validation evaluation criterion for the inversion results. For a reliable estimation, we impose a residual threshold: topographies generated with $\mathcal{R}$ below this threshold are deemed acceptable, whereas those that exceed it are discarded.
In the setup of this study, the distribution of $\omega$ during the validation process follows a normal distribution with parameters:
\begin{equation}
    \omega \sim \mathcal{N}(\mu = 5.0,\ \sigma = 2.0),
\end{equation}
where $\mu$ denotes the mean and $\sigma$ denotes the standard deviation of the sampling distribution.
To reduce computational cost, the number of validation runs is set to 30 in this study.

\subsection{Evaluation Metrics}

The $h,\hat h\in\mathbb{R}^{H\times W}$ denote the ground truth and the generated topography. For one sample of test set, the MSE, MAE, and SSIM are calculated as follows:
\begin{align}
\mathrm{MSE}(h,\hat{h}) &= \frac{1}{HW}\sum_{i=1}^{H}\sum_{j=1}^{W}\big(h_{ij}-\hat{h}_{ij}\big)^2,\\
\mathrm{MAE}(h,\hat{h}) &= \frac{1}{HW}\sum_{i=1}^{H}\sum_{j=1}^{W}\big|h_{ij}-\hat{h}_{ij}\big|,\\
\mathrm{SSIM}(h,\hat{h}) &=
\frac{(2\mu_{h}\mu_{\hat{h}} + C_1)(2\sigma_{h\hat{h}} + C_2)}
{(\mu_{h}^2 + \mu_{\hat{h}}^2 + C_1)(\sigma_{h}^2 + \sigma_{\hat{h}}^2 + C_2)},
\end{align}
where, $\mu_h,\mu_{\hat h},\sigma_h^2,\sigma_{\hat h}^2,\sigma_{h\hat h}$ are statistics values. In our experiments, $C_1=10^{-4}$ and $C_2=9\times10^{-4}$.

\section{Results and Discussion}
Based on the preceding definitions, we conducted a detailed analysis and comparison of model performance on two datasets.  The model's hyperparameters are listed in the table below. For inference, we evaluated three different sampling strategies to investigate their impact on the quality of generated data. Finally, we carried out a solver-guided calibration experiment, where the solver not only computed the steady-state topography generated by DiffTopo but also output the corresponding residuals. These residuals were then used to adjust the guide weight $\omega$. The single topography data sets used were randomly divided into training and testing subsets in a ratio of \(8{:}2\),
with a total of \(2{,}000\).

\subsection{The generation and solver posterior of SMT}

\begin{figure}
    \centering
    \includegraphics[width=1\linewidth]{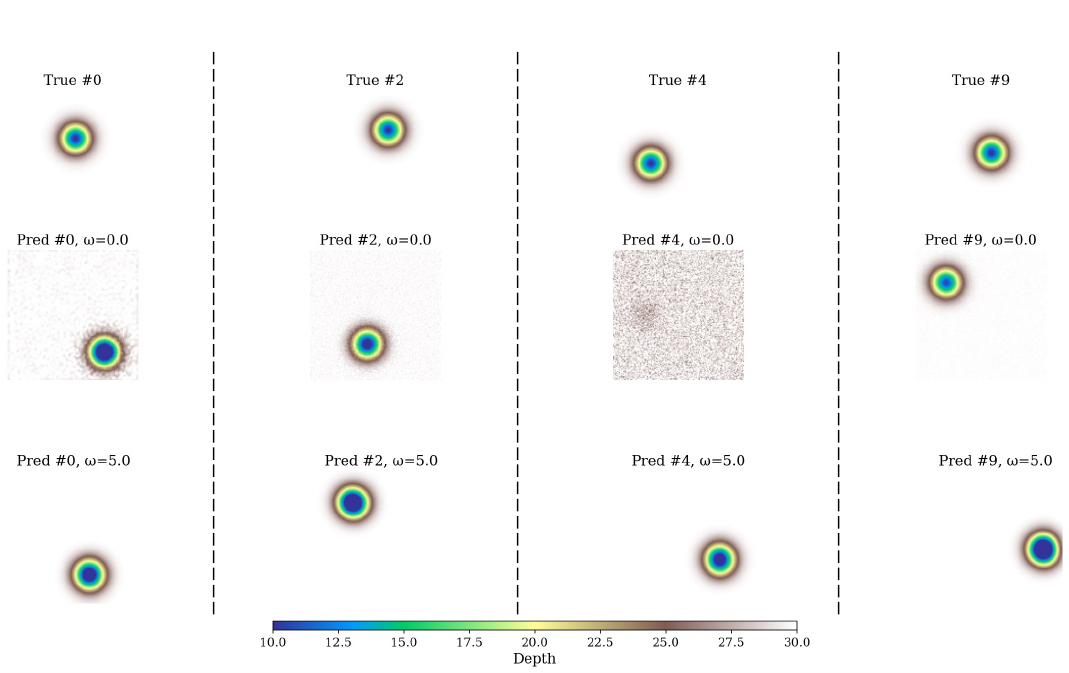}
    \caption{Sampling results using DPM with different guidance weights. The first row shows the ground truth topography, the second and third rows present results generated with guidance weight $\omega=0$ and $\omega=5$, respectively. The symbol ``\#'' denotes the indices of the test dataloader samples shown in each subplot.}
    \label{fig:smt_guidence}
\end{figure}

The performance comparison of three different samplers is shown in Table~\ref{tab:smt_table}. DPM++ achieves the highest SSIM score of 0.75, although it requires around 5 seconds for generation, lower than Heun. This indicates a strong similarity between the generated and ground-truth topography, as illustrated in Figure~\ref{fig:smt_guidence}. However, since the generative model only learns a probabilistic approximation within the data set distribution, the specific topography generated may vary from sample to sample with the different guidance weights. The second row in Figure~\ref{fig:smt_guidence} shows topographies generated randomly with a guidance weight of $\omega=0$. As observed, sample order 4 exhibits poor generation quality with strong noise artifacts. In contrast, third-row samples display topographies that closely resemble the ground truth, typically forming seamount-like structures. This phenomenon indicates that the guidance weight has an impact on the quality of the generated topographies. Lower $\omega$ encourages diversity, but may cause blur.
The higher $\omega$ improves sharpness while potentially sacrificing sample diversity. 
As shown in Appendix~\ref{fig:smt_train_dpm}, we visualize both the DDPM forward diffusion process and the DPM reverse sampling trajectory. This provides a clear depiction of how an individual sample is gradually corrupted with noise during training and denoised step-by-step during inference sampling.

\begin{figure}
    \centering
    \includegraphics[width=1\linewidth]{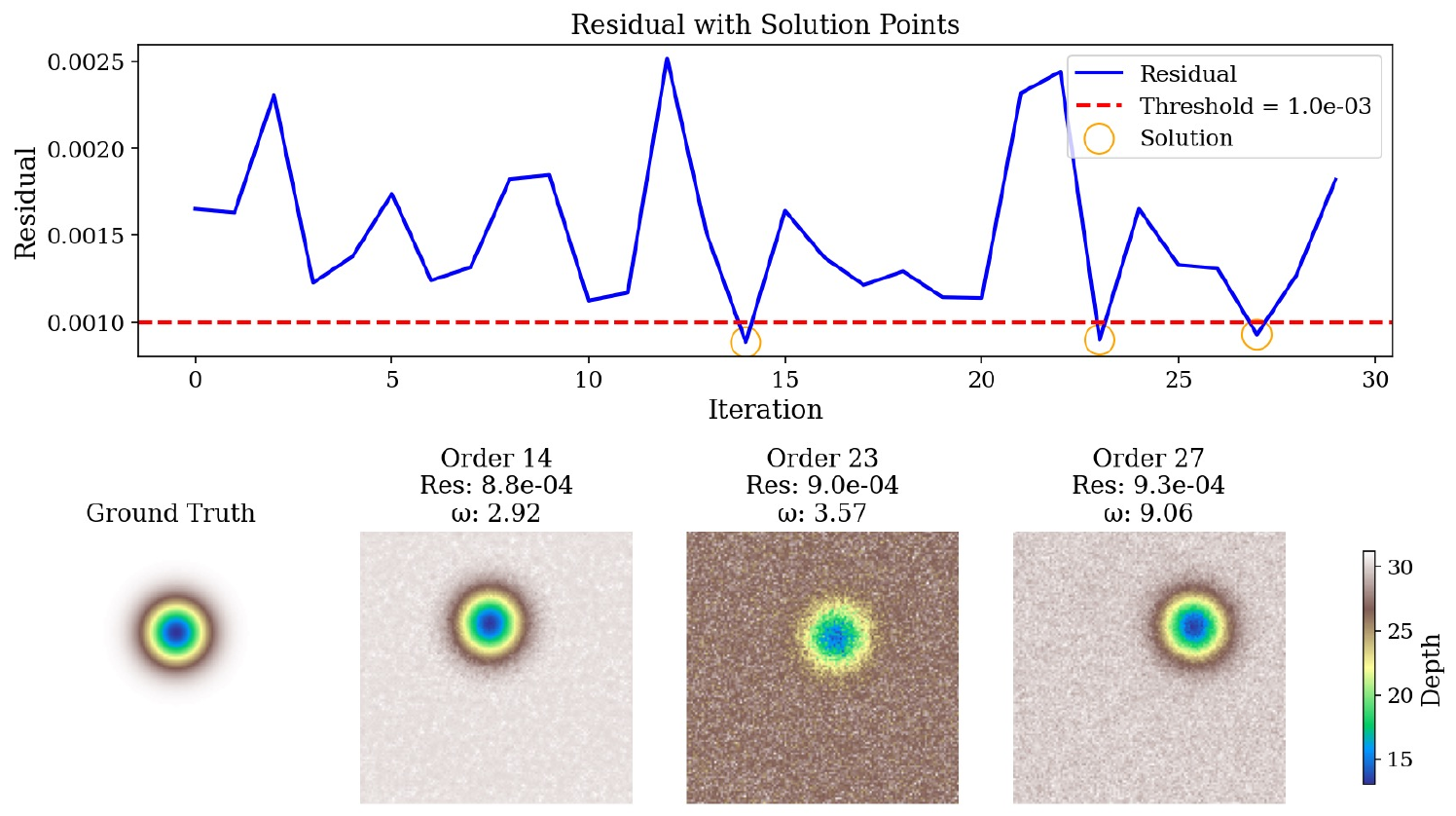}
    \caption{Posterior evaluation process using the solver on the SMT dataset.}
    \label{fig:smt_process_posterior}
\end{figure}

\begin{figure}
    \centering
    \includegraphics[width=0.8\linewidth]{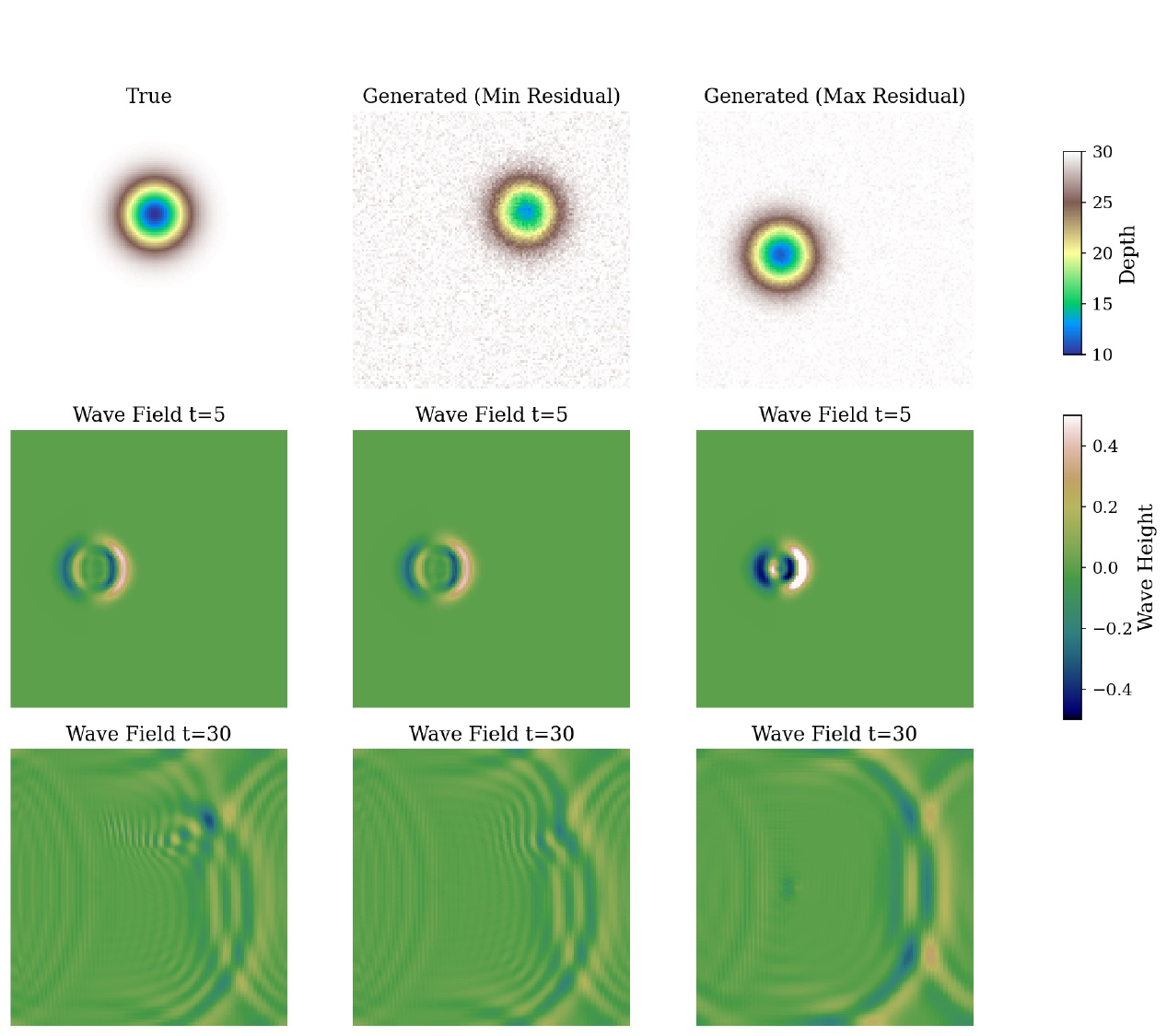}
    \caption{Comparison of the best and worst topography generated in SMT, along with their corresponding $\eta^{valid}$ in the steps of 5 and 30.}
    \label{fig:slover_smt}
\end{figure}

\begin{table}[ht]
    \centering
    \caption{Performance comparison on the SMT test dataset across different schedulers during inference in the NVIDIA 4090 with the guidance weight  $\omega=0$ during sampling. }
    \begin{tabular}{c c c c c c}
        \toprule
        \textbf{Sampling Schedulers (steps)} & \textbf{MAE \( \downarrow \)} & \textbf{MSE \( \downarrow \)} & \textbf{SSIM \( \uparrow \)} & \textbf{Gen Time (s) \( \downarrow \)} \\
        \midrule
        \multirow{1}{*}{DDPM (1000)} & 1.57 $\pm$0.11 & 8.14 $\pm$0.15 & 0.6 $\pm$0.09 & 17.2 \\
        \multirow{1}{*}{Heun (25)} &2.6$\pm$ 0.62 & 16.9$\pm$0.06  & 0.12 $\pm$0.064 & 0.4 \\
        \multirow{1}{*}{DPM++ (25)} & 1.7 $\pm$ 0.24 & 18.57 $\pm$1.66 & 0.75$\pm$0.10 & 5.1 \\
        
        \bottomrule
    \end{tabular}
    \label{tab:smt_table}
\end{table}
As illustrated in the Figure~\ref{fig:smt_process_posterior},  it can be seen that three generations met the setting threshold of $1e^{-3}$, as indicated by the yellow circles. The 14th generation yielded the most accurate result, exhibiting a shape and position that was highly consistent with the ground truth. To further demonstrate the necessity of the threshold and investigate the distinctions between the observations and the validated results, although the result generated with the lowest residual contains some noise, its location closely matches the true seamount in Figure~\ref{fig:slover_smt}. After recalculating the wave field using the solver, the simulated output also aligns well with the observed data.  In contrast, the sample with the highest residual already deviates from the observation at \( t = 5 \), the wave has passed over the part of the seamount, and the discrepancy increases as the wave evolves in \( t = 30 \). This shows that DiffTopo can produce reliable topography estimates when validated through the solver. Moreover, the residual serves as a quantitative confidence indicator: the smaller the residual, the more trustworthy the reconstruction. The threshold acts as a hyperparameter value that is too low to hinder meeting the criterion, particularly for small point-source amplitudes, whereas values that are too high can yield inaccurate estimates.


\subsection{The generation performance of TanT}

\begin{figure}[h]
    \centering
    \includegraphics[width=1\linewidth]{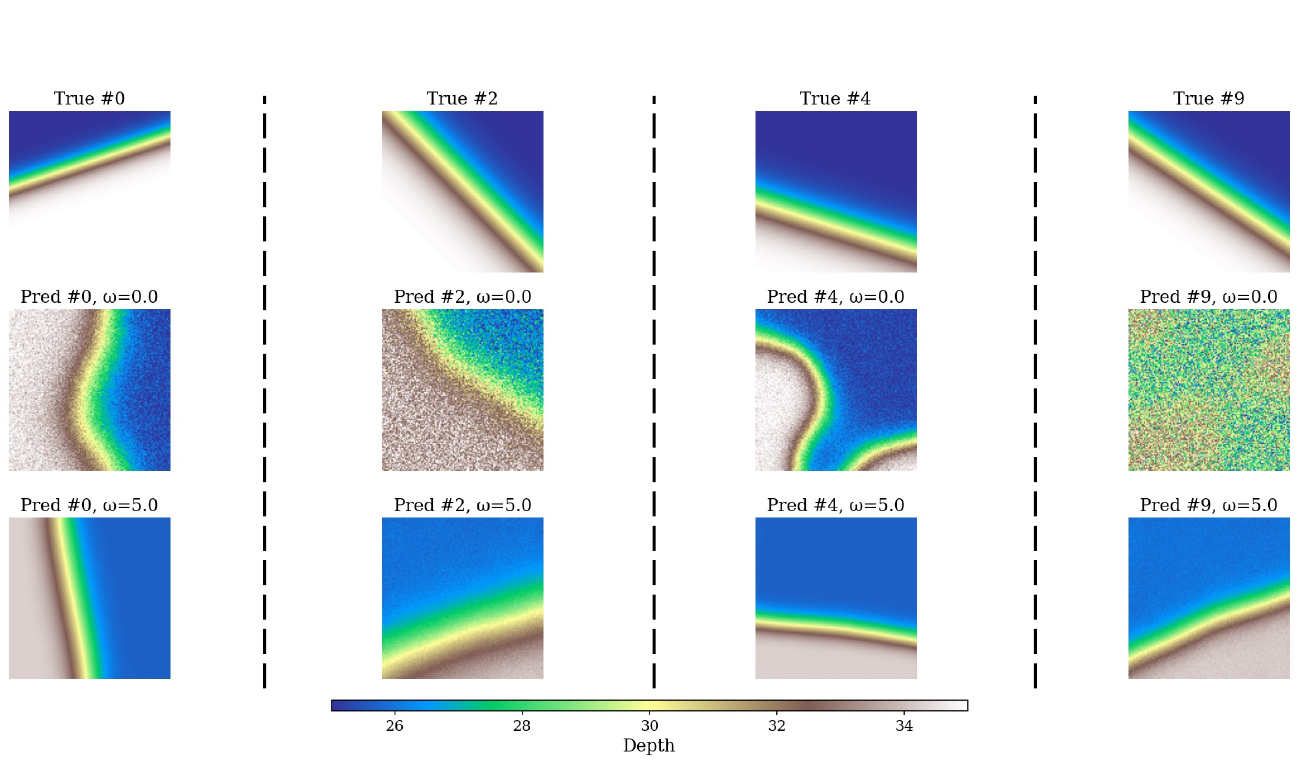}
    \caption{Sampling results using DPM with different guidance weights in TanT dataset.}
    \label{fig:DPM_sample_TanT}
\end{figure}

As shown in the Figure~\ref{fig:DPM_sample_TanT}, the guidance weight $w$ results in generation performance in the same way as in SMT. For lower values of
$\omega$, the generated results, as shown in the second row, appear more blurred. From the third row, it can be seen that increasing $w$ produces results that are markedly closer to the ground truth, both featuring a $\tanh$-like topography, although the direction is not exact. The results generated by $w = 0$ are shown in Table~\ref{tab:tanT_table}. As shown in Table~\ref{tab:tanT_table}, DPM++ achieves the lowest MAE ($3.20\pm0.69$) and MSE ($12.59\pm0.12$) on the TanT dataset with only 25 sampling steps, demonstrating superior numerical accuracy and efficiency compared to both DDPM and Heun. Although DDPM attains the highest SSIM ($0.42\pm0.015$), indicating better preservation of large-scale structures, its pointwise errors remain larger and it requires 1000 steps, leading to significantly higher computational cost. Heun, while using the same number of steps as DPM++, has both higher MAE and lower SSIM, suggesting that its second-order stochastic integration may be less effective for the ridge-like topographic features of TanT. Overall, these results highlight that DPM++ offers the best trade-off between accuracy and efficiency for this intermediate-difficulty terrain, whereas DDPM may still be preferable when structural similarity is prioritized over numerical fidelity.

\begin{figure}[h]
    \centering
    \includegraphics[width=1\linewidth]{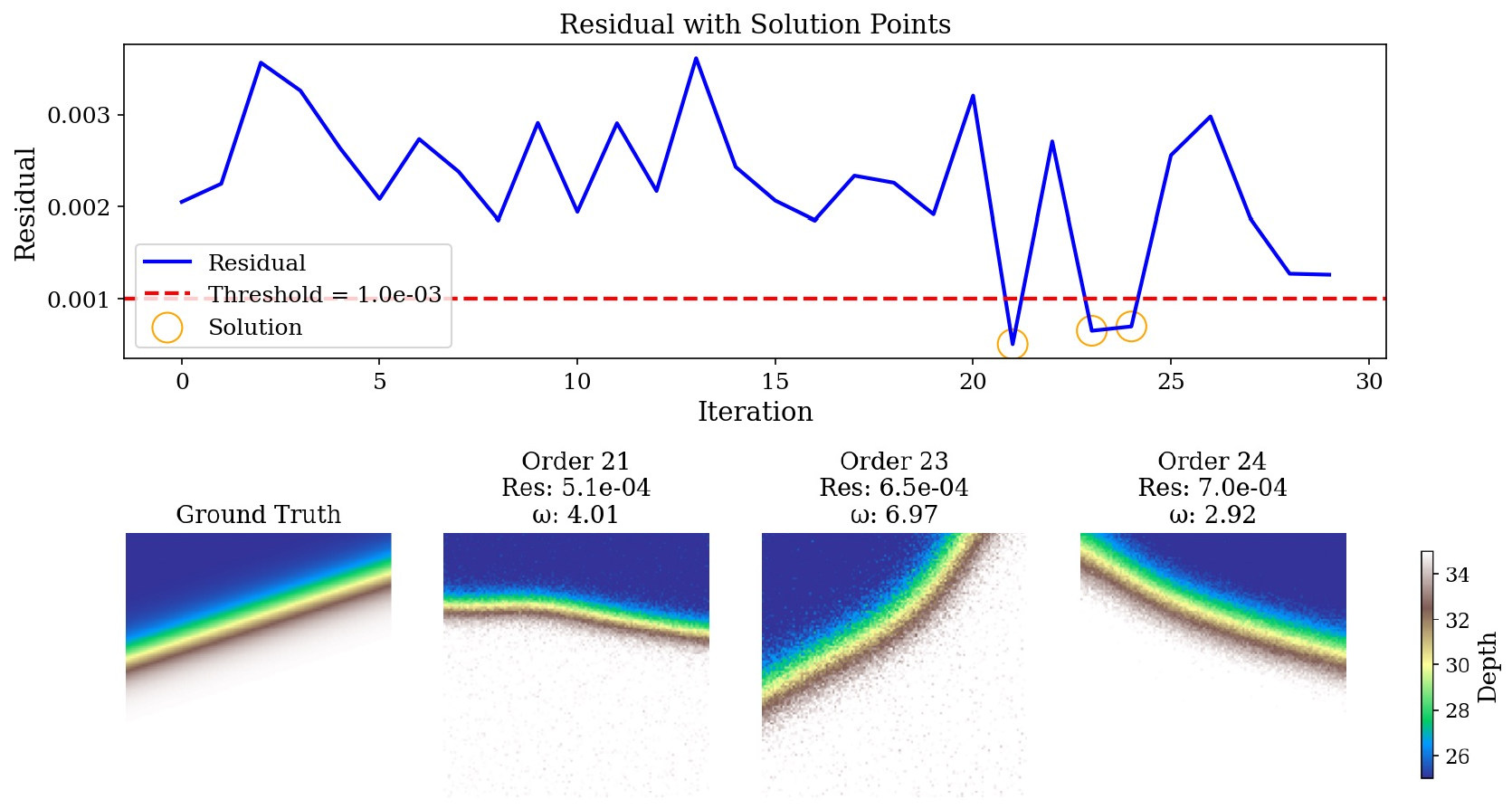}
    \caption{Posterior evaluation process using the solver on the TanT dataset.}
    \label{fig:valid_Tanh_Process}
\end{figure}

\begin{figure}[h]
    \centering
    \includegraphics[width=0.8\linewidth]{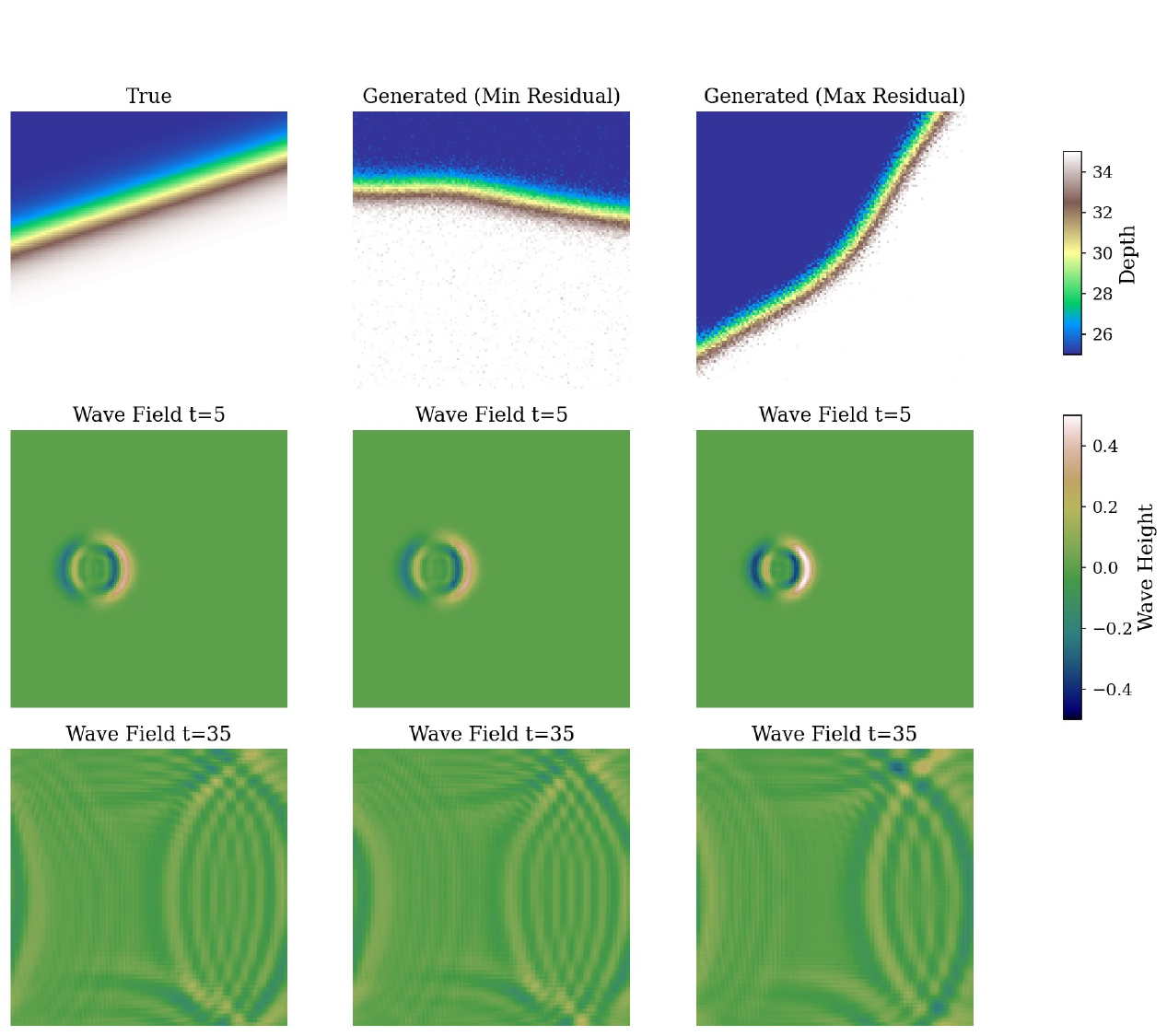}
    \caption{Comparison of the best and worst topography generated in TanT, along with their corresponding $\eta^{valid}$ in the steps of 5 and 35.}
    \label{fig:residual_show_tanh}
\end{figure}

The validation process is illustrated in Figure~\ref{fig:valid_Tanh_Process}, where, with a threshold set at $1\times 10^{-3}$, three feasible solutions were successfully obtained that meet both the solver's constraints and the threshold criterion. We compared the cases with the maximum and minimum residuals, as shown in Figure~\ref{fig:residual_show_tanh}. The comparison reveals that, although the topographies differ in shape, the generated topographies by DiffTopo are physically consistent with the target, indicating that the model has successfully learned the underlying distribution and satisfies the underdetermined nature of the inverse problem. In contrast, the worst-performing topography, despite having the correct orientation, results in the largest residual after solving, underscoring the necessity of the validation process.

\begin{table}[ht]
    \centering
    \caption{Performance comparison on the TanT test dataset across different schedulers during inference in the NVIDIA 4090 with the guidance weight  $\omega=0$ during sampling.}
    \begin{tabular}{c c c c}
        \toprule
        \textbf{Sampling Schedulers (steps)} & \textbf{MAE \( \downarrow \)} & \textbf{MSE \( \downarrow \)} & \textbf{SSIM \( \uparrow \)} \\
        \midrule
        \multirow{1}{*}{DDPM (1000)} & 3.66 $\pm$0.25 & 19.0 $\pm$0.022 & 0.42 $\pm$0.015 \\
        \multirow{1}{*}{Heun (25)} & 4.52$\pm$1.27 & 16.9$\pm$0.026 & 0.12 $\pm$0.09 \\
        \multirow{1}{*}{DPM++ (25)} & 3.2 $\pm$0.69 & 12.59 $\pm$0.12 & 0.34$\pm$0.014 \\
        \bottomrule
    \end{tabular}
    \label{tab:tanT_table}
\end{table}

\subsection{The generation performance of MMT}

\begin{table}[H]
    \centering
    \caption{Performance comparison on the MMT test dataset across different schedulers during inference in the NVIDIA 4090 with the guidance weight  $\omega=0$ during sampling.}
    \begin{tabular}{c c c c}
        \toprule
        \textbf{Sampling Schedulers (steps)} & \textbf{MAE \( \downarrow \)} & \textbf{MSE \( \downarrow \)} & \textbf{SSIM \( \uparrow \)} \\
        \midrule
        \multirow{1}{*}{DDPM (1000)} & 2.7 $\pm$0.13 & 11.1 $\pm$0.035& 0.23$\pm$0.05 \\
        \multirow{1}{*}{Heun (25)} & 2.4$\pm$0.18 & 10.2$\pm$0.057 & 0.16 $\pm$0.04 \\
        \multirow{1}{*}{DPM++ (25)} & 1.7 $\pm$0.10 & 10.57 $\pm$0.066 & 0.30$\pm$0.02 \\
        \bottomrule
    \end{tabular}
    \label{tab:mmt_table}
\end{table}

\begin{figure}[H]
    \centering
    \includegraphics[width=1\linewidth]{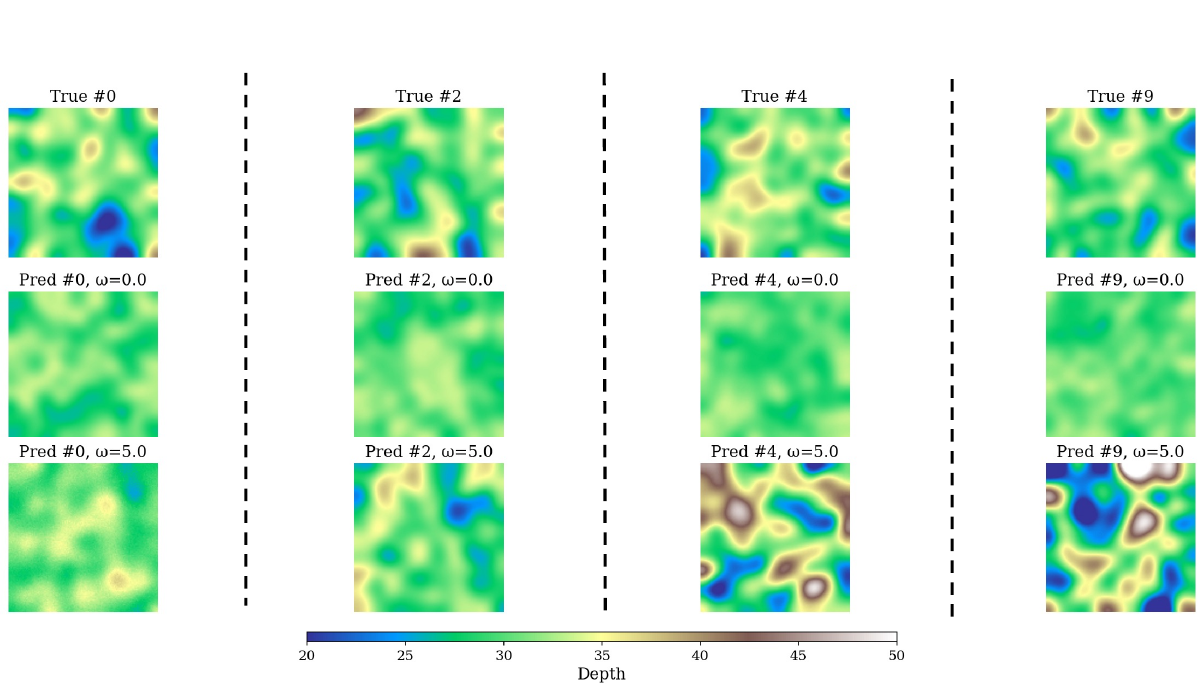}
    \caption{Sampling results using DPM with different guidance weights in MMT.}
    \label{fig:mmt_sample}
\end{figure}

As shown in Figure~\ref{fig:mmt_sample}, the generated topography results are shown in the second and third columns for $w=0$ and $w=5$, respectively. When $w=0$, the generated features are barely discernible, whereas with $w=5$, the generated topography exhibits a closer correspondence to the ground truth, although there are discrepancies. Compared with SMT and TanT, the quality of MMT generation is noticeably inferior. The worse performance is attributed to the increased difficulty of the MMT task: the target topographies exhibit higher stochasticity and heterogeneity, which amplifies the ill-posedness of the inverse problem and exceeds the current capacity of DiffTopo to learning the complex samples. Consistently, the training reconstruction loss in MMT is approximately one order of magnitude larger than in SMT and TanT, indicating both harder optimization and a poorer fit rather than a transient training instability.

As reported in Table~\ref{tab:mmt_table}, DPM++ again delivers the best overall performance in the MMT data set, achieving the lowest MAE ($1.70\pm0.10$) and MSE ($10.57\pm0.066$) while also obtaining the highest SSIM ($0.30\pm0.02$) among the three samplers, despite using only 25 sampling steps. This is particularly notable given that MMT represents a more challenging multi-peak topography with stronger nonlinearity and higher spatial variability compared to SMT and TanT. DDPM with 1000 sampling steps produces lower SSIM ($0.23\pm0.05$), highlighting its inefficiency in complex terrain scenarios. Heun performs slightly better than DDPM in MAE and MSE but remains inferior to DPM++ in all metrics, suggesting that higher-order deterministic solvers are particularly advantageous for accurately reconstructing intricate multi-peak bathymetries.

\begin{figure}[H]
    \centering
    \includegraphics[width=1\linewidth]{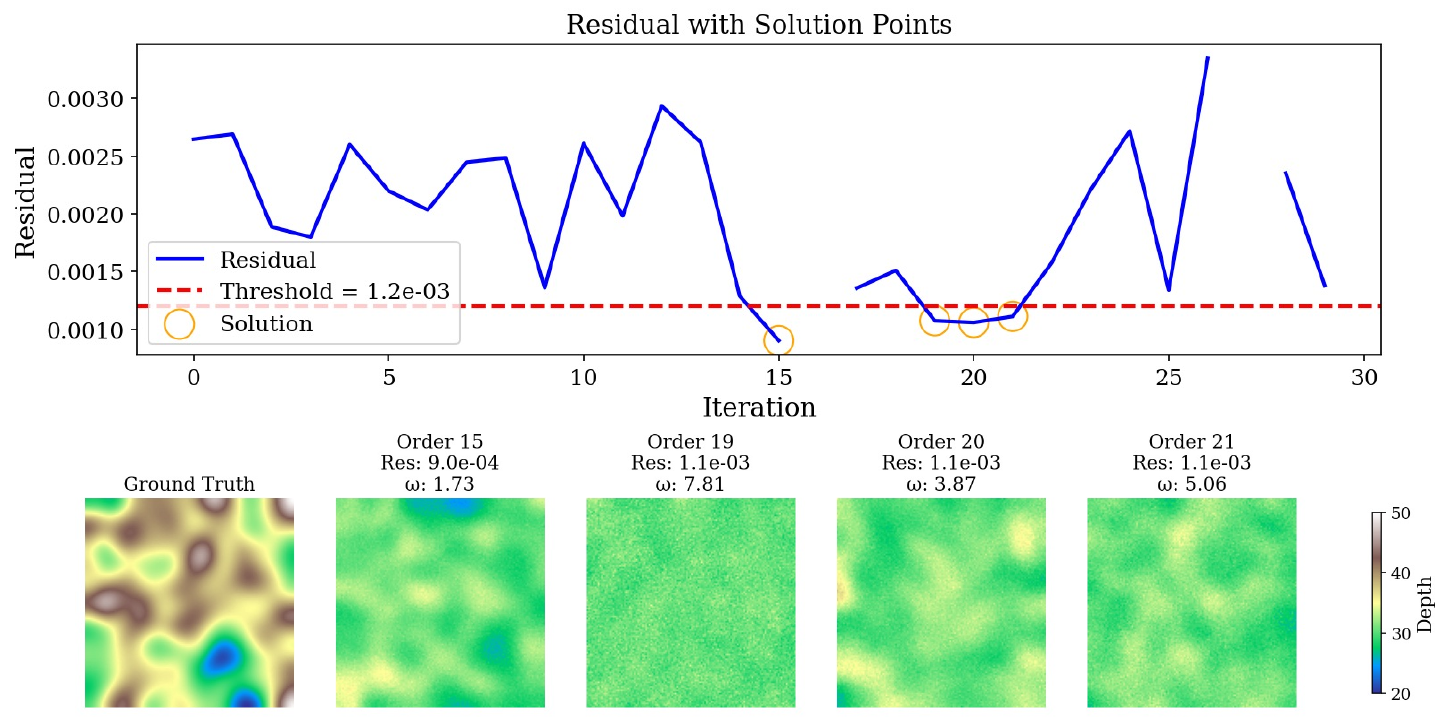}
    \caption{Posterior evaluation process using the solver on the MMT dataset.}
    \label{fig:mmt_valid}
\end{figure}

\begin{figure}[H]
    \centering
    \includegraphics[width=0.8\linewidth]{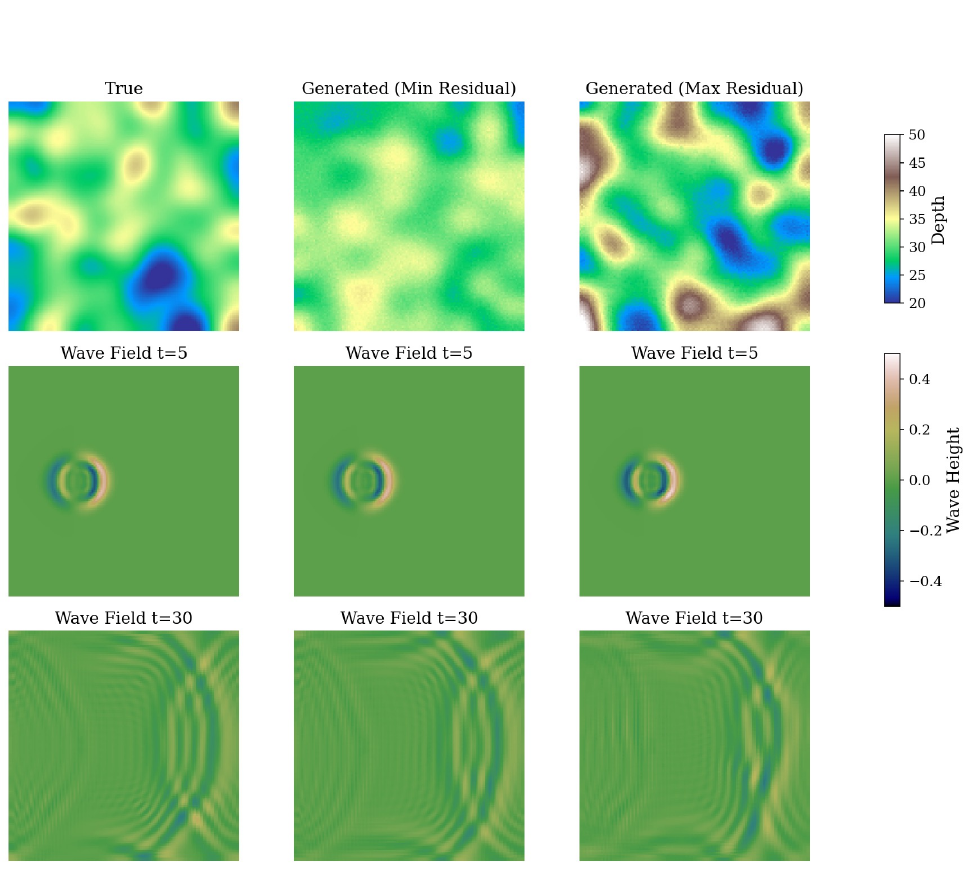}
    \caption{Comparison of the best and worst topography generated in MMT, along with their corresponding $\eta^{valid}$ in the steps of 5 and 30.}
    \label{fig:resiual_mmt}
\end{figure}

Figure~\ref{fig:mmt_valid} illustrates the validation process using the solver. When the threshold was set to $1\times10^{-3}$, feasible solutions were rarely obtained. By relaxing the criterion to $1.2\times10^{-3}$, four feasible solutions are within the setting threshold. The observed discontinuities in the curve arise from instances in which the quality of the generated topography causes the solver to produce NaN values, indicating that the generated results are not adequate to satisfy the solver’s numerical requirements. We present a representative sampling case showing the ground-truth topography and the corresponding wave fields obtained by the solver in Figure~\ref{fig:resiual_mmt}. In the first row, the topography associated with the smallest residual closely matches the ground truth, and in the second row, differences in the wave field are already observable at \(t = 5\). In \(t = 30\) (third row), the discrepancies remain relatively minor. In contrast, for the case with the largest residual, the wave field exhibits pronounced distortion at \(t = 30\), demonstrating that our validation procedure is effective in improving the reliability of the generated results.

\section{Conclusions}

This paper introduces DiffTopo, an innovative diffusion-based framework that incorporates solver feedback to achieve controllable and physically consistent terrain generation. We systematically evaluated the approach in three representative topography configurations and compared the performance of three sampling strategies: DDPM, Heun, and DPM++.

Experimental results demonstrate that DPM++ consistently outperforms the other samplers, achieving the best balance between generation quality and computational efficiency. From the dataset perspective, SMT proved to be the most tractable, largely due to its close resemblance to Gaussian distributions, which aligns well with the theoretical underpinnings of diffusion models. In contrast, MMT posed substantial challenges due to its highly irregular multi-peak structures and strong coupling between wave propagation and bathymetric complexity, while TanT exhibited intermediate difficulty.

These findings highlight both the promise and limitations of diffusion-based approaches for inverse topography problems. While DiffTopo performs well in  Gaussian-like topographies, its applicability to complex, high-variance topographies remains constrained by the inherent ill-posedness of the inverse problem. Future research will focus on advancing terrain generation for challenging MMT cases, potentially through hybrid modeling strategies that integrate physical priors, adaptive sampling schemes, or multiresolution representations. Furthermore, exploring cross-domain generalization and coupling with uncertainty quantification techniques may further enhance the robustness and interpretability of diffusion-based topography reconstruction.

\newpage
\bibliographystyle{plainnat}  
\bibliography{main}

\section{Appendix}

\subsection{Setting of hyper-parameters}

\begin{table}[h]
\centering
\caption{Architecture of UNet.}
\begin{tabular}{lccc}
\toprule
Stage & Layer & Input shape & Output shape \\
\midrule
Cond. projection & Conv2D & $(B,48,H,W)$ & $(B,1,H,W)$ \\
Encoder 1 & ResidualConvBlock & $(B,2,H,W)$ & $(B,64,H,W)$ \\
Down 1 & MaxPool2D & $(B,64,H,W)$ & $(B,64,H/2,W/2)$ \\
Encoder 2 & ResidualConvBlock & $(B,64,H/2,W/2)$ & $(B,128,H/2,W/2)$ \\
Down 2 & MaxPool2D & $(B,128,H/2,W/2)$ & $(B,128,H/4,W/4)$ \\
Encoder 3 & ResidualConvBlock & $(B,128,H/4,W/4)$ & $(B,256,H/4,W/4)$ \\
Down 3 & MaxPool2D & $(B,256,H/4,W/4)$ & $(B,256,H/8,W/8)$ \\
Bottleneck & ResidualConvBlock & $(B,256,H/8,W/8)$ & $(B,512,H/8,W/8)$ \\
Up 3 & ConvTranspose2D & $(B,512,H/8,W/8)$ & $(B,256,H/4,W/4)$ \\
Decoder 3 & ResidualConvBlock & $(B,512,H/4,W/4)$ & $(B,256,H/4,W/4)$ \\
Up 2 & ConvTranspose2D & $(B,256,H/4,W/4)$ & $(B,128,H/2,W/2)$ \\
Decoder 2 & ResidualConvBlock & $(B,256,H/2,W/2)$ & $(B,128,H/2,W/2)$ \\
Up 1 & ConvTranspose2D & $(B,128,H/2,W/2)$ & $(B,64,H,W)$ \\
Decoder 1 & ResidualConvBlock & $(B,128,H,W)$ & $(B,64,H,W)$ \\
Output & Conv2D & $(B,64,H,W)$ & $(B,1,H,W)$ \\
Residual add & Element-wise sum & $(B,1,H,W)$ & $(B,1,H,W)$ \\
\bottomrule
\end{tabular}
\end{table}

\begin{table}[h]
\centering
\caption{Training hyperparameters.}
\begin{tabular}{lc}
\toprule
Parameter & Value \\
\midrule
Optimizer & Adam\\
Learning rate & $1\times10^{-3}$ \\
Batch size & 40 \\
Training epochs & 1000 \\
DDPM timesteps & 1000 \\
Gradient accumulation & None \\
Normalization & Mean-Std  \\
Paras of UNET & 14.8M\\

\bottomrule
\end{tabular}
\end{table}
\paragraph{Data preprocessing} 
The input wave fields $\mathbf{\eta} \in \mathbb{R}^{T\times H\times W}$ and the topography fields $\hat{h} \in \mathbb{R}^{H\times W}$
were normalized using the mean and standard deviation calculated throughout the data set from the training split.
\[
\tilde{\mathbf{\eta}} = \frac{\mathbf{\eta} - \mu_\eta}{\sigma_\eta}, \quad 
\tilde{h} = \frac{h- \mu_h}{\sigma_h}.
\]
Normalization statistics are saved and reused for testing.

\subsection{Training loss}

\begin{figure}[htbp]
    \centering
    \includegraphics[width=0.9\linewidth]{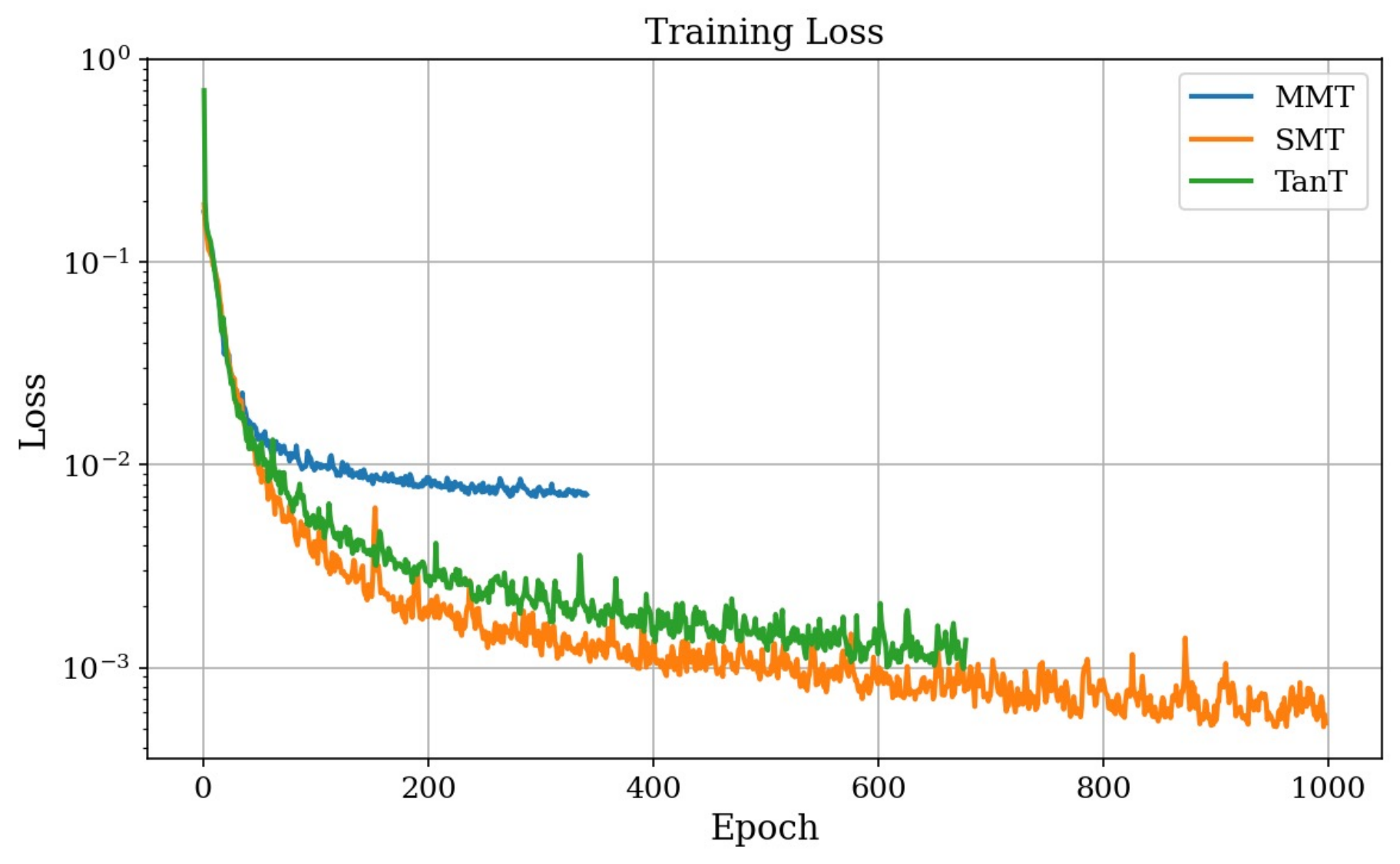}
    \caption{Training loss curves on the three datasets, with early stopping patience set to 50 epochs.  }
    \label{fig: The compare loss}
\end{figure}
\begin{figure}
    \centering
    \includegraphics[width=1\linewidth]{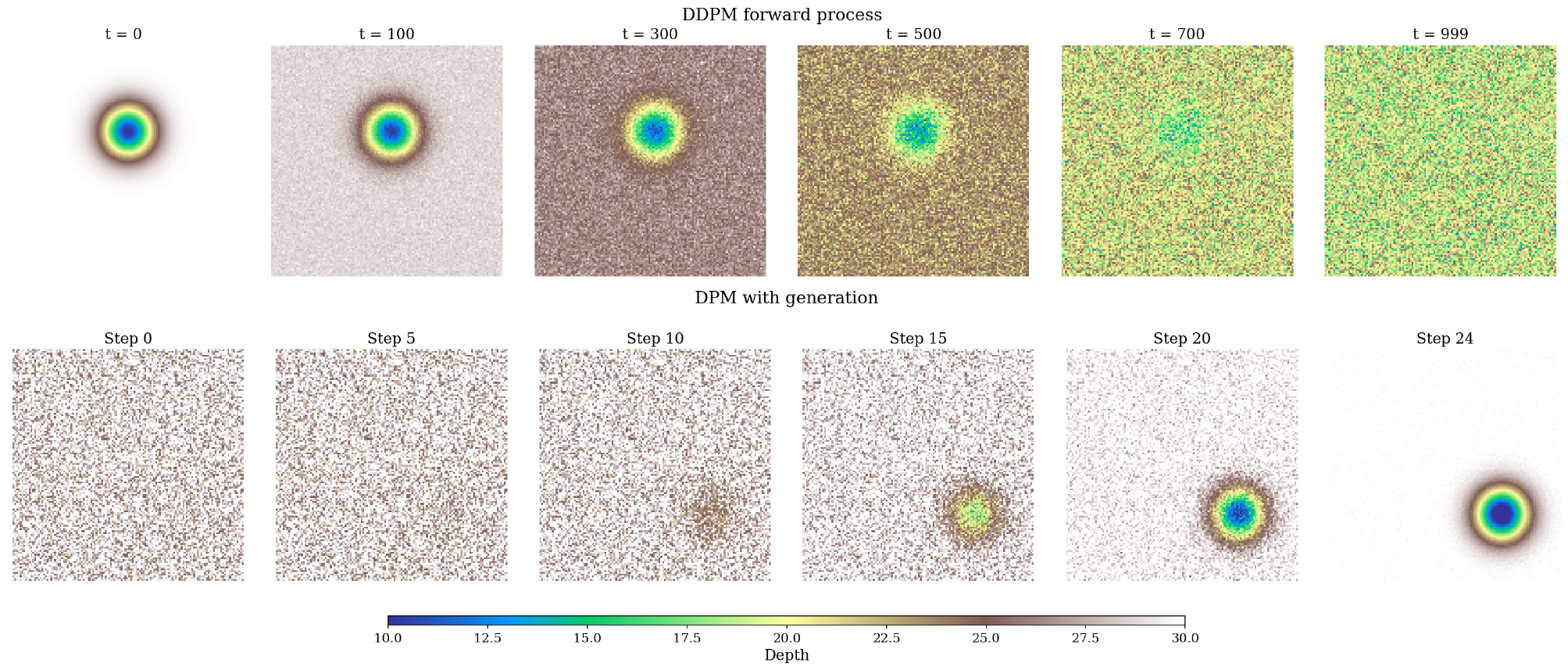}
    \caption{Comparison between the DDPM forward diffusion process and the reverse sampling trajectory of DPM. The top row represents the training-time noise injection at different diffusion timesteps t, while the bottom row illustrates the denoising steps during inference, indexed by step.}
    \label{fig:smt_train_dpm}
\end{figure}

Figure~\ref{fig: The compare loss} compares the training loss curves for the SMT, TanT, and MMT datasets using the same hyperparameters. The SMT and TanT datasets, being relatively simple, converge more rapidly to a loss level near $10^{-3}$, while the more complex MMT data set reaches only around $10^{-2}$. Since diffusion models are computationally intensive and typically stabilize around $10^{-2}$, we did not further optimize the training model for this study. The representative training and sampling process in the SMT is shown in Figure~\ref{fig:smt_train_dpm}.

\subsection{Principles of Classifier-Free Guidance: a score View}

\paragraph{Score consistency under \texorpdfstring{$\epsilon$}{epsilon}.}
In the variance-preserving forward process
\begin{equation}
q(x_t \mid x_0)=\mathcal{N}\!\big(\sqrt{\bar\alpha_t}\,x_0,\,(1-\bar\alpha_t)\mathbf I\big), 
\qquad \bar\alpha_t=\textstyle\prod_{s=1}^t\alpha_s,\ \alpha_s=1-\beta_s,
\end{equation}
Let
\begin{equation}
s_\star(x_t)\triangleq\nabla_{x_t}\log q(x_t),
\qquad 
s_\star(x_t\!\mid\!c)\triangleq\nabla_{x_t}\log q(x_t\!\mid\!c)
\end{equation}
denote the unconditional and conditional scores. Denoising score matching yields the identities
\begin{equation}
\mathbb E[\epsilon\mid x_t] = -\sqrt{1-\bar\alpha_t}\,s_\star(x_t),
\qquad
\mathbb E[\epsilon\mid x_t,c] = -\sqrt{1-\bar\alpha_t}\,s_\star(x_t\!\mid\!c).
\end{equation}
Training the network \(\epsilon_\theta\) to predict \(\epsilon\) induces the scores
\begin{equation}
s_\theta(x_t) \;\triangleq\; -\frac{1}{\sqrt{1-\bar\alpha_t}}\,\epsilon_\theta(x_t,\varnothing),
\qquad
s_\theta(x_t\!\mid\!c) \;\triangleq\; -\frac{1}{\sqrt{1-\bar\alpha_t}}\,\epsilon_\theta(x_t,c),
\end{equation}
which approach \(s_\star(x_t)\) and \(s_\star(x_t\!\mid\!c)\) at optimum. Bayes' rule gives the following.
\begin{equation}
\log q(x_t\!\mid\!c)=\log q(c\!\mid\!x_t)+\log q(x_t)-\log q(c).
\end{equation}
Taking \(\nabla_{x_t}\) yields
\begin{equation}
s_\star(x_t\!\mid\!c)-s_\star(x_t)=\nabla_{x_t}\log q(c\!\mid\!x_t).
\end{equation}
We therefore define the conditional signal as
\begin{equation}
\Delta s_\star(x_t;c)\;\triangleq\; s_\star(x_t\!\mid\!c)-s_\star(x_t)
\;=\; \nabla_{x_t}\log q(c\!\mid\!x_t),
\end{equation}
i.e., exactly the posterior gradient used by classifier guidance. Replacing the true scores by model scores gives
\begin{equation}
\Delta s_\theta(x_t;c)\;\triangleq\; s_\theta(x_t\!\mid\!c)-s_\theta(x_t)
\;\approx\; \nabla_{x_t}\log q_\theta(c\!\mid\!x_t).
\end{equation}

\paragraph{CFG as classifier guidance in score space.}
At inference time, classifier-free guidance combines conditional and unconditional noise predictions as
\begin{equation}
\hat\epsilon_\theta(x_t,c)=(1+w)\,\epsilon_\theta(x_t,c)-w\,\epsilon_\theta(x_t,\varnothing),\qquad w>0,
\end{equation}
which corresponds to the guided score
\begin{align}
\hat s_\theta(x_t;c,w)
&= -\frac{1}{\sqrt{1-\bar\alpha_t}}\,\hat\epsilon_\theta(x_t,c) \\
&= s_\theta(x_t) \;+\; w\big(s_\theta(x_t\!\mid\!c)-s_\theta(x_t)\big) \\
&= s_\theta(x_t) \;+\; w\,\Delta s_\theta(x_t;c) \\
&= \boxed{\,s_\theta(x_t) + w\,\nabla_{x_t}\log q_\theta(c\!\mid\!x_t)\,}.
\end{align}
Thus, CFG performs classifier guidance without training an external classifier: it adds a weighted ``\(\nabla\) conditional signal'' to the unconditional score.

\paragraph{Mean shift of the DDPM reverse step.}
The mean of one DDPM reverse step with condition \(\tilde c\in\{c,\varnothing\}\) is
\begin{equation}
\mu_\theta(x_t,\tilde c)
=\frac{1}{\sqrt{\alpha_t}}
\left(x_t-\frac{1-\alpha_t}{\sqrt{1-\bar\alpha_t}}\,\epsilon_\theta(x_t,\tilde c)\right).
\end{equation}
Let \(\mu_u(x_t)\triangleq \mu_\theta(x_t,\varnothing)\). The CFG mean shift relative to the unconditional step is
\begin{align}
\mu_{\text{CFG}}(x_t,c;w)-\mu_u(x_t)
&= -\frac{1}{\sqrt{\alpha_t}}\frac{1-\alpha_t}{\sqrt{1-\bar\alpha_t}}\,
\big(\hat\epsilon_\theta(x_t,c)-\epsilon_\theta(x_t,\varnothing)\big) \\
&= -\frac{1+w}{\sqrt{\alpha_t}}\frac{1-\alpha_t}{\sqrt{1-\bar\alpha_t}}\,
\big(\epsilon_\theta(x_t,c)-\epsilon_\theta(x_t,\varnothing)\big) \\
&= \frac{1+w}{\sqrt{\alpha_t}}\,(1-\alpha_t)\,\Delta s_\theta(x_t;c),
\end{align}
since \(\epsilon_\theta=-\sqrt{1-\bar\alpha_t}\,s_\theta\). Hence each reverse step moves the mean along the conditional signal \(\Delta s_\theta\), with step size scaled by \((1+w)(1-\alpha_t)/\sqrt{\alpha_t}\).

\paragraph{Geometric decomposition and control.}
Let \(\hat s_\theta=s_\theta+w\Delta s_\theta\). Let \(\theta_t\) be the angle between \(\Delta s_\theta\) and \(s_\theta\). Decompose into components parallel and orthogonal to \(s_\theta\):
\begin{equation}
\hat s_\theta 
= \big(\|s_\theta\| + w\|\Delta s_\theta\|\cos\theta_t\big)\frac{s_\theta}{\|s_\theta\|}
\;+\; w\|\Delta s_\theta\|\sin\theta_t\,\mathbf u_\perp,
\end{equation}
with \(\mathbf u_\perp\) a unit vector orthogonal to \(s_\theta\). Increasing \(w\) (i) increases the component along \(s_\theta\) (stronger push toward high-density regions) and (ii) rotates the direction toward the conditional score when \(\theta_t\neq 0\), improving alignment with the posterior manifold defined by \(c\).


\end{document}